\documentclass[twocolumn,10pt]{revtex4}

\usepackage{amsmath}
\usepackage{graphicx}
\usepackage{array}
\usepackage{multirow}

\begin{document}

\title{Rebound-through transition of bright-bright solitons collision in two species condensates with repulsive interspecies interactions}
\author{Z.M. He, D.L. Wang$\footnote{%
E-mail: dlwang@xtu.edu.cn}$, J.W. Ding} \affiliation{$$Department
of Physics ${\&}$ Institute for Nanophysics and Rare-earth
Luminescence, Xiangtan University, Xiangtan 411105, China}
\date{\today }

\vskip .2cm
\begin{abstract}
 \vskip 0.5cm

 \noindent\textbf{Abstract:} We study the dynamical properties of bright-bright solitons in two species Bose-Einstein condensates
 with the repulsive interspecies interactions under the external harmonic potentials by using a variational
 approach combined with numerical simulation. It is found that the interactions between bright-bright solitons
 vary from repulsive to attractive interactions with the increasing of their separating distances. And the bright-bright solitons
 can be localized at equilibrium positions, different from the periodic oscillation of bright soliton in the single species condensates. Especially,
 a through-collision is newly observed from the bright-bright solitons collisions with the increasing of the initial velocity. The collisional
 type of bright-bright solitons, either rebound - or through -collision, depends on the modulation of the initial conditions. These results will
 be helpful for the experimental manipulating such solitons.
\vskip 0.3cm
 \noindent\textbf{PACS (2008)}: 05.45.Yv, 03.75.Lm, 03.75.Mn
 \vskip 0.3cm
 \noindent\textbf{Keywords:}        Bright-bright solitons ${\bullet}$ two species BECs ${\bullet}$ collisions
\end{abstract}

\maketitle

\vskip 2.5cm
 {\noindent \textbf{1. Introduction}}

\vskip .5 cm

Recently, matter wave soliton pairs of dark-dark solitons and
dark-bright solitons have been experimentally observed in two
species Bose-Einstein condensates (BECs) [1-7]. In the experiment,
an external magnetic gradient is applied to make the two species
BECs undergo phase separation [1, 2]. In the phase separation
regime, the bright soliton in one species can be stably trapped
inside a density dip of the dark soliton in the other species [3],
while it can not occur in the single species BECs. This means that
the lifetime of bright soliton in two-species BECs [3] is longer
than that of one in the single species BECs [8, 9]. This
long-lived bright solitons may open possibilities for future
applications in coherent atomic optics, atom interferometry, and
atom transport.

\vskip 0.2cm

 The possibility of creating soliton
pairs in two species BECs has stimulated considerable theoretical
interest in their existence conditions, stability, interactions,
collisions, and so on [10-34]. Based on the mean-field theory, the
solitons properties in the two species BECs are usually described
by the coupled Gross-Pitaevskii (GP) equation, which is similar to
the coupled nonlinear Schr$\ddot{\textrm{o}}$dinger equation
(NLSE) used in nonlinear optics. Some literatures showed that in
the absence of the external potentials, the interactions between
the bright-bright solitons (BBSs) are mainly determined by the
interspecies interactions of two species BECs [10-12]. In this
case, for the attractive (repulsive) interspecies interaction,
there is an attractive (repulsive) potential between the BBSs, and
thus they exhibited always attraction [10, 11] (repelsion [12])
each other. Taking into account the external harmonic potential,
we here explore the dynamical properties of BBSs in two species
BECs. It is found that the interactions between BBSs vary from
repulsive to attractive interactions with the increasing of their
separating distances, even for repulsive interspecies
interactions. And the BBSs can be localized at the equilibrium
position, different from the periodic oscillation of bright
soliton in the single species condensates.

\vskip 0.2cm

In addition, the interspecies interactions of two species BECs
have important effect on the collisional properties of BBSs
[13-17]. For the attractive interactions, a through-collision (TC)
occurs between the BBSs [13, 14]. Such solitons pass through each
other at the collision point, and the TC is independent of the
initial conditions, such as the separating distances and
velocities [14]. For the repulsive interspecies interactions,
there exists a regime of elastic particle collisions [15], where
there occurs a rebound-collision (RC) with a momentum exchange
between the BBSs. It is worthwhile pointing out that, in nonlinear
optics, the collisional type of BBSs can be controlled by their
initial velocities [17].

\vskip 0.2cm

To our knowledge, there is little report on the effect of the
initial velocity and separating distance on the collisional
properties of BBSs in two species BECs with the repulsive
interspecies interactions. Therefore, considering the repulsive
interspecies interactions and external harmonic potentials, we
explore the collisions between BBSs with different initial
velocities and separating distance. It is found that a TC is newly
observed from the BBSs collisions with the increasing of the
initial velocity. And the collisional type of BBSs, either TC or
RC, depends on the modulation of the initial conditions.

\vskip 0.5cm

{\noindent \textbf{2.} \textbf{Model}}

 \vskip 0.3cm

We consider that the two species BECs are trapped in the harmonic
potentials
[30-34]$\emph{V}^{(\emph{i})}_{\emph{ext}}(\emph{X},\emph{Y},\emph{Z})=\emph{M}_{\emph{i}}[\omega^{2}_{\bot(\emph{i})}(\emph{Y}^{2}+\emph{Z}^{2})+\omega^{2}_{\emph{i}}\emph{X}^{2}]/2$.
Here, $\omega_{\bot(\emph{i})}$ and $\omega_{\emph{i}}$ are the
radial and transverse trapping frequencies with $\emph{i}=1, 2$,
and $\emph{M}_{\emph{i}}$ is the atomic mass of the
$\emph{i}^{th}$ species. If $\omega_{\bot(\emph{i})}\gg
\omega_{\emph{i}}$ , it is reasonable to reduce the GP equation to
a coupled one-dimensional NLSE
\begin{displaymath}
i\hbar\frac{\partial\Psi_{1}(X,T)}{\partial
T}=[-\frac{\hbar^{2}}{2M_{1}}\frac{\partial^{2}}{\partial X^{2}}+
2\hbar\omega_{\bot (1)}a_{1}N_{1}|\Psi_{1}|^{2}
\end{displaymath}
\begin{equation}
               +\frac{\hbar\omega_{\bot (1)}a_{12}M_{1}N_{1}}{M}|\Psi_{2}|^{2}+\frac{M_{1}\omega^{2}_{1}
           X^{2}}{2}]\Psi_{1},
\end{equation}

\begin{displaymath}
i\hbar\frac{\partial\Psi_{2}(X,T)}{\partial
T}=[-\frac{\hbar^{2}}{2M_{2}}\frac{\partial^{2}}{\partial X^{2}}+
\frac{2\hbar\omega_{\bot (1)}M_{1}a_{2}N_{2}}{M_{2}}|\Psi_{2}|^{2}
\end{displaymath}
\begin{equation}
               +\frac{\hbar\omega_{\bot (1)}M_{1}a_{12}N_{2}}{M}|\Psi_{1}|^{2}+\frac{M_{2}\omega^{2}_{2}
           X^{2}}{2}]\Psi_{2},
\end{equation}
where $N_{i} (\emph{i}=1,2)$ is the atoms number of the $i^{th}$
species; $a_{i}$ and $a_{12}$ are the intraspecies and
interspecies scattering length (SL), respectively; and $
1/M=1/M_{1}+1/M_{2}$. Subsequently, we introduce some
dimensionless variables $X=a_{\bot}x$, $T=2t/\omega_{\bot(1)}$,
and $\psi_{i}=a_{\bot}\Psi_{i} $, with
$a_{\bot}=\sqrt{\hbar/M_{1}\omega_{\bot(1)}}$, so that Eqs. (1)
and (2) are reduced into
\begin{equation}
[i\frac{\partial}{\partial t}+\frac{\partial^{2}}{\partial
x^{2}}+g_{11}|\psi_{1}|^{2}-g_{12}|\psi_{2}|^{2}-\frac{\omega^{2}_{1}x^{2}}{\omega^{2}_{\bot(1)}}]\psi_{1}=0,
\end{equation}
\begin{equation}
[i\frac{\partial}{\partial t}+\frac{\partial^{2}}{\partial
x^{2}}+\varepsilon g_{22}|\psi_{2}|^{2}-\varepsilon
g_{21}|\psi_{1}|^{2}-\frac{\omega^{2}_{2}x^{2}}{\varepsilon
\omega^{2}_{\bot(1)}}]\psi_{2}=0,
\end{equation}
where $\varepsilon=M_{1}/M_{2}$, $g_{ii}=-4a_{i}N_{i}/a_{\bot}$,
$g_{12}=2a_{12}M_{1}N_{1}/(Ma_{\bot})$, and
$g_{21}=2a_{12}M_{2}N_{2}/(Ma_{\bot})$. We here consider the two
species BECs composed of $^{7}$Li and $^{39}$K atoms which is
accessible for experiments, and the species one (two) is $^{7}$Li
($^{39}$K) condensate. Based on the currently experimental
conditions, the radial trapping frequencies are chosen as
$\omega_{\bot(1)}=\omega_{\bot(2)}=2\pi\times 100$ Hz, so that the
time and space units correspond to 6.4 ms and 5.4 $\mu m$ in
reality, respectively. We also choose the atom numbers
$N_{1}=N_{2}=2000$, the intraspecies SLs $a_{1}=-25 $ $a_{B}$ and
$a_{2}=-139 $ $a_{B}$, respectively, and interspecies SL
$a_{12}=-42.4  $ $a_{B}$. Here, $a_{B}$ is Bohr radius.
 \vskip 0.5cm

{\noindent \textbf{3. }\textbf{ Variational approach}}

 \vskip 0.5cm

 When the interspecies interactions were repulsive, the BBSs
 exhibited repelsive each other [12]. It means that there is a repulsive
 potential between the BBSs. In order to obtain this effective potential,
 we solve Eqs. (3) and (4) by using a variational approach [10].
 We here propose that $\omega_{1}=\omega_{2}=0$, in this case,
  the ansatz solitons solutions of Eqs. (3) and (4) are chosen as
\begin{equation}
\psi_{j}(x,t)=\frac{a(t)\exp[i\eta_{j}(t)(x-x_{j}(t))+i\mu_{j}(t)]}{\cosh[\kappa
a(t)(x-x_{j}(t))]},
\end{equation}
where  $\kappa$  is constant; $a$, $x_{j}$, $ \eta_{j}$, and
$\mu_{j}$ are the functions of time $t$ with $j=1, 2$. So, the
Lagrangian density $ \Gamma$ can be given by
\begin{displaymath}
\Gamma=\sum^{2}_{j=1}[\frac{i}{2}(\bar{\psi_{j}}\frac{\partial
\psi_{j}}{\partial t}-\psi_{j}\frac{\partial
\bar{\psi_{j}}}{\partial t})-|\frac{\partial \psi_{j}}{\partial
x}|^{2}
\end{displaymath}
\begin{equation}
+\frac{g_{jj}|\psi|^{4}}{2}]-g_{12}|\psi_{1}|^{2}|\psi_{2}|^{2}.
\end{equation}
Here, the overbar denotes the complex conjugate. Substituting Eq.
(5) into Eq. (6) and then integrating the result over $x$ from
$-\infty$ to $\infty$, we obtain the Lagrangian
\begin{displaymath}
L=\sum^{2}_{j=1}[2\frac{a}{\kappa}(\eta_{j}\frac{\partial
x_{j}}{\partial t}-\eta^{2}_{j}-\frac{\partial \mu_{j}}{\partial
t})+\frac{2a^{3}(g_{jj}-\kappa^{2})}{3\kappa}]
\end{displaymath}
\begin{equation}
-\frac{4g_{12}a^{3}A(x_{2}-x_{1})}{\kappa},
\end{equation}
where
\begin{equation}
 A=\frac{{\kappa a(x_{2}-x_{1})\cosh[\kappa
a(x_{2}-x_{1})]-\sinh[\kappa a(x_{2}-x_{1})]}}{\sinh^{3}[\kappa
a(x_{2}-x_{1})]}.
\end{equation}
The action is defined as $I=\int^{t2}_{t1}L d t$ . We can get a
set of equations for the ansatz parameters from the least-action
principle $\delta I=0$ . They are given by
\begin{equation}
\delta I / \delta \mu_{j}=0\rightarrow a=\textrm{constant},
\end{equation}
\begin{equation}
\delta I / \delta\eta_{j}=0\rightarrow \partial x_{j}/\partial
t=2\eta_{j},
\end{equation}
\begin{equation}
\delta I /\delta x_{j}=0\rightarrow \partial \eta_{j}/\partial
t=-2g_{12}a^{2}\partial A/ \partial x_{j},
\end{equation}
\begin{displaymath}
\delta I /\delta a=0\rightarrow \partial \mu_{j}/\partial
t=\eta_{j}\partial x_{j}/\partial
t-\eta^{2}_{j}
\end{displaymath}
\begin{equation}
+2a^{2}(g_{jj}-\kappa^{2})-4g_{12}a^{2}A/3,
\end{equation}
where $j=1, 2$. From the Eqs. (10) and (11), we get
\begin{equation}
\partial_{t}(a^{2}\frac{\partial x_{1}}{\partial t}+a^{2}\frac{\partial x_{2}}{\partial
t})=-4g_{12}a^{4}(\partial_{x_{1}}+\partial_{x_{2}})A=0,
\end{equation}
and Newton's motion equation
\begin{equation}
a^{2}\frac{\partial^{2} x_{j}}{\partial
t^{2}}=-4g_{12}a^{4}\frac{\partial A}{\partial x_{j}},
\end{equation}
where $j=1,2$. Accordingly, we obtain
\begin{equation}
a^{2}\frac{\partial x_{1}}{\partial t}+a^{2}\frac{\partial
x_{2}}{\partial t}=P,
\end{equation}
\begin{equation}
\frac{1}{2}a^{2}\frac{\partial x^{2}_{1}}{\partial
t}+\frac{1}{2}a^{2}\frac{\partial x^{2}_{2}}{\partial
t}+4g_{12}a^{4}A=E,
\end{equation}
where $P$ and $E$ are integral constants. If we take $a^{2}$ as
the mass of the soliton, equations (15) and (16) represent the
conservation of momentum and energy of the solitons, respectively.
In Eq. (16), we can define the effective potential $V_{eff} (q)$
[10]
\begin{equation}
V_{eff}(q)=4g_{12}a^{4}[\frac{{\kappa aq\cosh(\kappa
aq)-\sinh(\kappa aq)}}{\sinh^{3}(\kappa aq)}].
\end{equation}
Here $q=x_{2}-x_{1}$ represents the separating distances of two
solitons. For a better understanding this effective potential
between the BBSs, we plot the effective potential varying with the
separating distance in Fig. 1. One can see that with the
increasing value of separating distance q, the effective potential
$V_{eff} (q)$ decreases. And when $q$ increases a defined value,
the effective potential vanishes. This manifests that the force
between the BBSs is a short-range force.

\begin{figure}[tbp]
\includegraphics[width=3.4in]{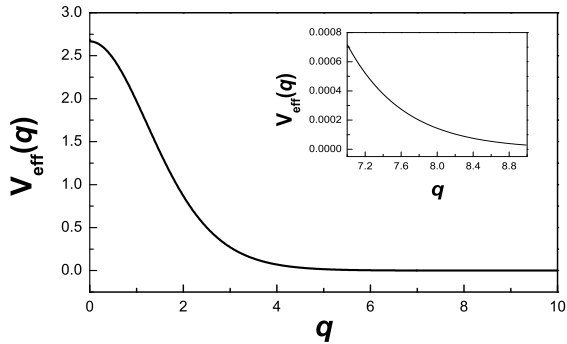}
\caption{ Variations of the effective potential with the
separating distance. The parameters used are $a=1.0$,
$g_{12}=2.0$, and $\kappa=0.882$. }
\end{figure}

\vskip 0.5cm

{\noindent \textbf{4. }\textbf{ The interaction forces between two
solitons}}

 \vskip 0.5cm

To obtain interaction forces between the BBSs, we here numerically
simulate the Eq. (2) by the Crank-Nicolson method [34]. The
initial conditions are chosen as [15]
\begin{equation}
\psi_{j}(x,0)=\frac{b_{j}\exp(2ic_{j}x)}{\cosh[b_{j}(x-x_{j})]},
\end{equation}
where $b_{j}$, $x_{j}$, and $c_{j}$ are constant with $j=1, 2$. To
get the range of the interspecies interaction force, we first
consider that the transverse trapping frequencies
$\omega_{1}=\omega_{2}=0$. The corresponding bright solitons
positions varying with the time $t$ are shown in Fig. 2. One can
see from Fig. 2(a) that at the initial time, the bright soliton of
species one is at $x_{1}=3.00$, and the bright soliton of (the
other) species two is at  $x_{2}=-3.00$. The initial velocities of
solitons are set as  $c_{1}=c_{2}=0$. With the time going on, the
bright soliton of species one moves along the positive direction
of \emph{x}-axis, and the other one moves along the negative
direction of \emph{x}-axis. This indicates that there exists a
repulsive force between the bright solitons. While the initial
positions are displaced at $x_{1}=-x_{2}=4.20$ and the initial
velocities are still set as $c_{1}=c_{2}=0$ [see Fig. 2(b)], the
position of each soliton keeps unchanged with the time going on.
It means that the repulsive force between two solitons vanishes.
In this case, we can obtain that the range of this repulsive force
is the separating distance $q \leq 8.40$. In addition, from the
inset of Fig. 1, it makes sure that when the separating distance
$q=8.40$, the  $V_{eff} (q) \approx 10^{-4}\approx 0$. So the
numerical result is in good agreement with the result of the
variational approach.
\begin{figure}[tbp]
\includegraphics[width=3.4in]{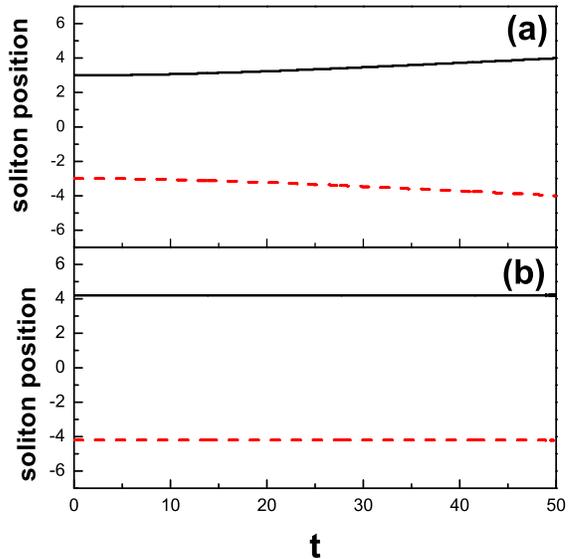}
\caption{ (Color online) Evolvement of the solitons positions in
two species BECs with the repulsive interspecies interactions. The
initial positions of two solitons are (a) $x_{1}=2.00$,
$x_{2}=-2.00$, and (b) $x_{1}=4.20$, $x_{2}=-4.20$, respectively.
The other parameters used are $b_{1}=1.0$, $b_{2}=1.0$, $c_{1}=0$,
$c_{2}=0$, $g_{11}=2.0$, $g_{22}=2.0$, $g_{12}= 2.0$,
$\omega_{1}=0$, and $\omega_{2}= 0$. Solid and dash lines
correspond to the species one and two, respectively. }
\end{figure}

\vskip0.3cm

We then choose the transverse trapping frequencies
$\omega_{1}=1.4\pi$ Hz and $\omega_{2}=0.6\pi$ Hz to explore the
interaction forces between the BBSs. Figure 3 shows the
corresponding solitons positions varying with the time. In this
case, the initial velocities of two solitons are still set as
$c_{1}=c_{2}=0$. One can see from Fig. 3(a) that at the initial
time, the bright soliton of species one is at  $x_{1}=2.00$, and
the bright soliton of species two is at  $x_{2}=-2.00$. With the
time going on, the bright soliton of species one moves along the
positive direction of \emph{x}-axis, and the other one moves along
the negative direction of \emph{x}-axis. This indicates that the
interaction between two soliton is repulsive. While the initial
positions of two solitons are displaced at  $x_{1}=-x_{2}=4.00$
[see Fig. 3(b)], it is interesting to see that the bright soliton
of species one moves along the negative direction of
\emph{x}-axis, and the other one moves along the positive
direction of \emph{x}-axis. This indicates that the interaction
between two solitons is attractive. We here can conclude that the
interactions between the BBSs vary from repulsive to attractive
with the increasing of the separating distance. In fact, when two
solitons are trapped in the harmonic potentials, each soliton
undergoes two forces which are from the external potentials and
interspecies interactions. For convenience, we here defined
 $F_{1(i)}$  and $F_{2(i)}$ $(i=1, 2)$  represent the forces coming from the
external potentials and interspecies interactions, respectively.
That is $F_{1(i)}=-d V_{trap}(x_{i})/ d
x_{i}=-2\omega^{2}_{x(1)}x_{i}/ \omega^{2}_{\bot(1)}$ and
$F_{2(i)}=-\partial V_{eff}(q)/ \partial x_{i}$  with $i=1, 2$.
When the two bright solitons are at $x_{1}=2.00 $ and
$x_{2}=-2.00$, respectively, [such as in Fig. 3(a)], we have
$F_{1(1)}=-F_{1(2)}\approx -0.00800$ and
$F_{2(1)}=-F_{2(2)}\approx 0.13428$. Duo to that
$|F_{2}|>|F_{1}|$,   two solitons undergo repulsive interactions
each other. While the two bright solitons are at $x_{1}=4.00$ and
$x_{2}=-4.00$, respectively, [such as in Fig. 3(b)], there are
$F_{1(1)}=-F_{1(2)}\approx -0.01600$ and
$F_{2(1)}=-F_{2(2)}\approx 0.00030$, so that $|F_{2}|<|F_{1}|$ and
thus two solitons attract each other. That is to say, the
interactions between two solitons, either attractive or repulsive,
depend on the difference between the absolute value of $F_{1}$ and
$F_{2}$.

\begin{figure}[tbp]
\includegraphics[width=3.4in]{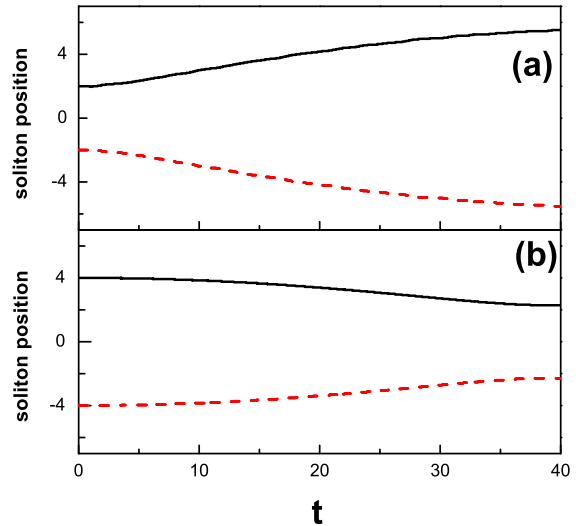}
\caption{ (Color online) Evolvement of the solitons positions in
two species BECs with the repulsive interspecies interactions and
external harmonic potentials. The initial positions of two
solitons are (a) $x_{1}=2.00$, $x_{2}=-2.00$, and (b)
$x_{1}=4.00$, $x_{2}=-4.00$, respectively. The parameters used are
$\omega_{1}=1.4\pi$ Hz and $\omega_{2}=0.6\pi$ Hz. The other
parameters used are the same as the figure 2. Solid and dash lines
correspond to the species one and two, respectively. }
\end{figure}
\vskip0.3cm

In addition, if the two bright solitons newly displace at
$x_{1}=2.84$  and $x_{2}=-2.84$, respectively, there are
$F_{1(1)}=-F_{1(2)}\approx -0.01136$ and
$F_{2(1)}=-F_{2(2)}\approx 0.01135$, so that
$|F_{2}|\approx|F_{1}|$. In this case, the propagating
characteristics of two bright solitons are shown in Fig. 4. One
sees that the width, amplitude, and position of each soliton keep
unchanged with time. Obviously, they are localized solitons,
different from the periodic oscillation of bright soliton in the
single species BECs [35,36]. Thanks to the two balance forces, the
lifetime of the bright solitons in two-species BECs at the
equilibrium positions should be longer than that of one in the
single BECs. These stable bright solitons may open possibilities
for future applications in coherent atomic optics.

\begin{figure}[tbp]
\includegraphics[width=3.4in]{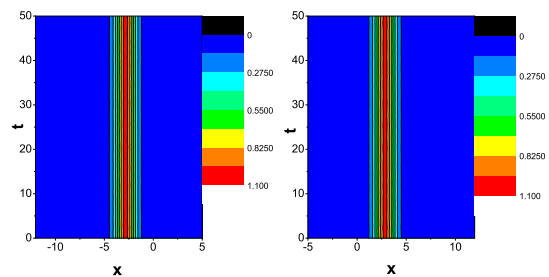}
\caption{ (Color online) The propagating characteristics of BBSs
in two species BECs with the repulsive interspecies interactions
and external harmonic potentials. The parameters used are
$x_{1}=2.84$, $x_{2}=-2.84$, $\omega_{1}=1.4\pi$ Hz and
$\omega_{2}=0.6\pi$ Hz. The other parameters are the same as the
figure 2. The right and left plots represent the density of the
species one and two, respectively. }
\end{figure}
\vskip0.3cm

From discussed above, we conclude that the interaction forces
between BBSs of two species BECs trapped in the harmonic
potentials exhibit either attraction or repulsion, which is
controlled by their initial separating distance. And the BBSs can
be localized at equilibrium positions of two equal forces.

\vskip 0.5cm

{\noindent \textbf{5. }\textbf{ The collision properties of BBSs}}

 \vskip 0.5cm

In order to explore the effect of initial velocity on the
collision properties of BBSs, we here consider the two bright
solitons are at the equilibrium positions (which are at
$x_{1}=2.84$ and $x_{2}=-2.84$, respectively). Figure 5 shows the
soliton positions as a function of the time with the different
initial velocity. When the initial velocities of two solitons are
$c_{1}=0.1$ and $c_{2}=-0.1$ [as shown in Fig. 5(a)],
respectively, the two solitons move towards each other and take
place a collision at $t\approx 5$. They do not pass through each
other at the collision point, so, this collision belongs to RC.
This phenomenon is similar with the report in Ref. [15]. While the
initial velocities of two solitons are increased to $c_{1}=0.7$
and $c_{2}=-0.7$ [see Fig. 5(b)], respectively. Interestingly, it
is observed that the two bright solitons pass through each other
at the time $t\approx 1.2$. This collision is TC. The followed TC
behavior can be observed at $t\approx 10$ and $t\approx18.5$. This
shows that the collision type of BBSs exhibits a transition from
RC to TC.
\vskip0.3cm
\begin{figure}[tbp]
\includegraphics[width=3.4in]{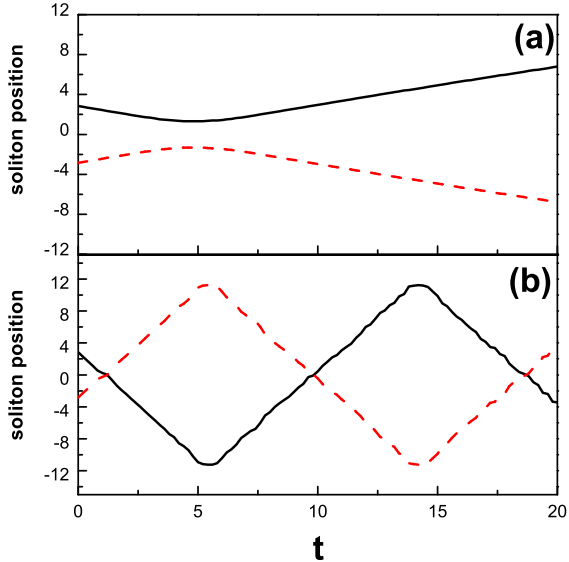}
\caption{ (Color online) Evolvement of the solitons positions with
the different initial velocities. The initial velocities of two
solitons are (a) $c_{1}=0.1$, $c_{2}=-0.1$, (b) $c_{1}=0.7$,
$c_{2}=-0.7$, respectively. The parameters used are $x_{1}=2.84$,
$x_{2}=2.84$, $\omega_{1}=1.4\pi$ Hz and $\omega_{2}=0.6\pi$ Hz.
The other parameters are the same as the figure 2. Solid and dash
lines correspond to the species one and two, respectively. }
\end{figure}

Subsequently, in order to get the effect of initial separating
distance on the collision properties of BBSs, we propose that the
initial velocities of two solitons are $c_{1}=0.8$ and
$c_{2}=-0.8$, respectively. The corresponding collisional
behaviors of BBSs with different initial separating distance are
plotted in Fig. 6. When two solitons are at $x_{1}=2.00$ and
$x_{2}=-2.00$, respectively [see Fig. 6(a)], one can see that the
collisions between two solitons are all TC. While the initial
positions of two solitons are newly displaced at $x_{1}=5.20$ and
$x_{2}=-5.20$ [as shown in Fig. 6(b)], respectively, it is
interesting to observe that the collisions between two solitons
are all RC.
\vskip0.3cm
\begin{figure}[tbp]
\includegraphics[width=3.4in]{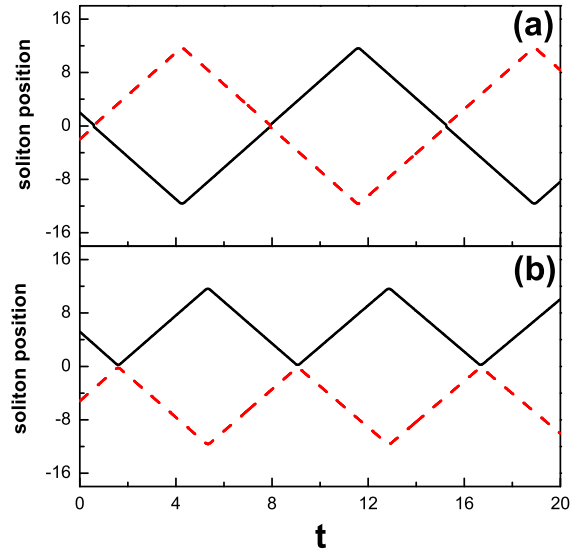}
\caption{ (Color online) Evolvement of the solitons positions in
two species BECs with the different initial separating distance.
The initial positions of two solitons are (a) $x_{1}=2.00$,
$x_{2}=-2.00$, (b) $x_{1}=5.20$, $x_{2}=-5.20$, respectively. The
parameters used are $b_{1}=2.0$, $b_{2}=2.0$, $c_{1}=0.8$,
$c_{2}=-0.8$, $\omega_{1}=1.4\pi$ Hz and $\omega_{2}=0.6\pi$ Hz.
The other parameters are the same as the figure 2. Solid and dash
lines correspond to the species one and two, respectively. }
\end{figure}

We here can conclude that the collisional type of BBSs, either TC
or RC, depends on their initial velocities and separating
distance. These results will be helpful for the experimental
manipulating such BBSs.

\vskip 0.3cm

{\noindent \textbf{6. }\textbf{Conclusion}}

 \vskip 0.3cm

In summary, we investigate the dynamical properties of BBSs in two
species BECs with the repulsive interspecies interaction under the
external harmonic potentials. Using the variational approach, we
obtain the effective repulsive potential inducing by the
interspecies interactions. By analyzing this effective potential,
we find that the repulsive force between the BBSs is a short-range
force. Then, we numerical simulate the dynamical properties of
BBSs in the coupled one-dimensional NLSE by the Crank-Nicolson
method. It is shown that the interactions between BBSs vary from
repulsive to attractive interaction with the increasing of their
separating distance. And the BBSs can be localized at equilibrium
positions. These stable solitons may open possibilities for future
applications in coherent atomic optics. Moreover, it is found that
the TC is newly observed from the BBSs collisions with the
increasing of their initial velocities. And the collisional type
of BBSs, either TC or RC, depends on the modulation of the initial
conditions.


\vskip 0.3cm \noindent \textbf{References}



\begin{thebibliography}{99}

\bibitem{1} C. Hamner, J.J. Chang, P. Engels, M.A. Hoefer, Phys. Rev. Lett. 106, 065302 (2011)

\bibitem{2} M.A. Hoefer, J.J. Chang, C. Hamner, P. Engels, Phys. Rev. A 84, 041605 (2011)

\bibitem{3} C. Becker, S. Stellmer, P. Soltan-Panahi, S. D$\ddot{\textrm{O}}$rscher, M. Baumert, Eva-Maria Richter, Jochen Kronj$\ddot{\textrm{a}}$ger, K. Bongs, K. Sengstock, Nat. Phys. 4, 496 (2008)

\bibitem{4}  G. Thalhammer, G. Barontini, L. DeSarlo, J. Catani, F. Minardi, M. Inguscio, Phys. Rev. Lett. 100, 210402 (2008)

\bibitem{5} Th. Best, S. Will, U. Schneider, L. Hackerm¨¹ller, D. van Oosten, I. Bloch, D.-S. L¨¹hmann, Phys. Rev. Lett. 102, 030408 (2009)

\bibitem{6} S.B. Papp, J.M. Pino, C.E. Wieman, Phys. Rev. Lett. 101, 040402 (2008)

\bibitem{7} G. Roati, M. Zaccanti, C. D'Errico, J. Catani, M. Modugno, A. Simoni, M. Inguscio, G. Modugno, Phys. Rev. Lett. 99, 010403 (2007)

\bibitem{8} K.E. Strecker, G.B. Partridge, A.G. Truscott, R.G. Hulet, Nature 417, 150 (2002)

\bibitem{9} L. Khaykovich, F. Schreck, G. Ferrari, T. Bourdel, J. Cubizolles, L.D. Carr, Y. castin, C. Salomon, Science 296, 1290 (2002)

\bibitem{10} H.Y. Yu, L.X. Pan, J.R. Yan, J.Q. Tang, J. Phys. B: At. Mol. Opt. Phys. 42, 0253019 (2009)

\bibitem{11} S.K. Adhikari, Phys. Lett. A 346, 179 (2005)

\bibitem{12} G. Csire, D. Schumayer, B. Apagyi, Phys. Rev. A 82, 063608 (2010)

\bibitem{13} X.F. Zhang, X.H. Hu, X.X. Liu, W.M. Liu, Phys. Rev. A 79, 033630 (2009)

\bibitem{14} L. Salasnich, B.A. Malomed, Phys. Rev. A 74, 053610 (2006)

\bibitem{15}D. Novoa, B.A. Malomed, H. Michinel, V.M. P$\acute{\textrm{e}}$rez-Garc$\acute{\textrm{i}}$a, Phys. Rev. Lett. 101, 144101 (2008)

\bibitem{16} V. M. P$\acute{\textrm{e}}$rez-Garc$\acute{\textrm{i}}$a, J. B. Beitia, Phys. Rev. A 72, 033620 (2005)

\bibitem{17}  J.K. Yang, Y. Tan, Phys. Rev. Lett. 85, 3624 (2000)

\bibitem{18} X.X. Liu, H.Pu, B. Xiong, W.M. Liu, J.B. Gong, Phys. Rev. A 79, 013423 (2009)

\bibitem{19} D.S. Wang, X.H. Hu, W.M. Liu, Phys. Rev. A 82, 023612 (2010)

\bibitem{20}U. Shrestha, J. Javanainen, J. Ruostekoski, Phys. Rev. Lett. 103, 190401 (2009)

\bibitem{21} K.J.H. Law, P.G. Kevrekidis, L.S. Tuckerman, Phys. Rev. Lett. 105, 160405 (2010)

\bibitem{22} Th. Busch, J. R. Anglin, Phys. Rev. Lett. 87, 010401 (2001)

\bibitem{23} P. $\ddot{\textrm{O}}$ hberg, L. Santos, Phys. Rev. Lett. 86, 2918 (2001)

\bibitem{24}  P. $\ddot{\textrm{O}}$hberg, L. Santos, J. Phys. B: At. Mol. Opt. Phys. 34, 4271 (2001)

\bibitem{25} M. Vijayajayanthi, T. Kanna, M. Lakshmanan, Phys. Rev. A 77, 013820 (2008)

\bibitem{26} L. Li, B.A. Malomed, D. Mihalache, W.M. Liu, Phys. Rev. E 73, 066610 (2006)

\bibitem{27} L. Li, Z.D. Li, B.A. Malomed, D. Mihalache, W.M. Liu, Phys. Rev. A 72, 033611 (2005)

\bibitem{28} V.A. Brazhnyi, V.V. Konotop, Phys. Rev. E 72, 026616 (2005)

\bibitem{29} H.E. Nistazakis, D.J. Frantzeskakis, P.G. Kevrekidis, B.A. Malomed, R. Carretero-Gonz¨¢lez, Phys. Rev. A 77, 033612 (2008)

\bibitem{30} A.I. Yakimenko, Y.A. Zaliznyak, V.M. Lashkin, Phys. Rev. A 79, 043629 (2009)

\bibitem{31} S.K. Adhikari, Phys. Lett. A 346, 179 (2005)

\bibitem{32} V.M. Lashkin, E.A. Ostrovskaya, A.S. Desyatnikov, Y.S. Kivshar, Phys. Rev. A 80, 013615 (2009)

\bibitem{33} L.C. Zhao, S.L. He, Phys. Lett. A 375, 3017 (2011)

\bibitem{34} K. Kasamatsu, M. Tsubota, Phys. Rev. A 74, 013617 (2006)

\bibitem{35} U. Al Khawaja, H.T.C. Stoof, R.G. Hulet, K.E. Strecker, G.B. Partridge, Phys. Rev. Lett. 89, 200404 (2002)

\bibitem{36} P.K. Shukla, D.D. Tskhakaya, Phys. Scripta 107, 259 (2004)


\end{thebibliography}
\end{document}